\documentclass[journal, 10pt]{IEEEtran}

\usepackage{multirow}
\usepackage[font=scriptsize,caption=false,labelsep=space]{subfig}
\usepackage{mathrsfs}
\usepackage{graphicx,float,wrapfig,epstopdf,amsmath}
\usepackage[]{algorithmicx}
\usepackage{algpseudocode,algorithm}
\usepackage{balance}
\usepackage{mathtools,lipsum}
\usepackage{amsmath}
\usepackage{amssymb}
\usepackage{amsthm}
\usepackage{graphicx}
\usepackage{epstopdf}
\usepackage{times}
\usepackage{textcomp,cite}

\usepackage{url}
\usepackage{hyperref}
\usepackage{comment}

\usepackage{multirow}
\usepackage{threeparttable}

\usepackage[table]{xcolor}
\definecolor{intnull}{RGB}{213,229,255}
\definecolor{inteins}{RGB}{128,179,255}
\definecolor{color1}{RGB}{199,209,232}
\definecolor{color2}{RGB}{230,231,233}

\usepackage{multirow}

\begin{document}
	
	\title{Deep Inverse Design of Reconfigurable Metasurfaces for Future Communications}
	\author{\IEEEauthorblockN{John A. Hodge, Kumar Vijay Mishra and Amir I. Zaghloul}
		\thanks{J. A. H. and A. I. Z. are with Bradley Department of Electrical and Computer Engineering, Virginia Tech, Blacksburg, VA 24061 USA. Email: \{jah70, amirz\}@vt.edu.} 
		\thanks{K. V. M. and A. I. Z. are with United States CCDC Army Research Laboratory, Adelphi, MD 20783 USA. E-mail: kumarvijay-mishra@uiowa.edu, amirz@vt.edu.}
		\thanks{J. A. H. acknowledges support from Northrop Grumman Mission Systems (NGMS), Baltimore, MD, for his thesis research. K. V. M. acknowledges support from the National Academies of Sciences, Engineering, and Medicine via Army Research Laboratory Harry Diamond Distinguished Postdoctoral Fellowship.}
	}
	\maketitle
	
	\begin{abstract}
		Reconfigurable intelligent surfaces (RIS) have recently received significant attention as building blocks for smart radio environments and adaptable wireless channels. By altering the space- and time-varying electromagnetic (EM) properties, the RIS transforms the inherently stochastic nature of the wireless environment into a programmable propagation channel. Conventionally, designing RIS to yield a desired EM response requires trial-and-error by iteratively investigating a large possibility of various geometries and materials through thousands of full-wave EM simulations. In this context, deep learning (DL) techniques are proving critical in reducing the computational cost and time of RIS inverse design. Instead of explicitly solving Maxwell’s equations, DL models learn physics-based relationships through supervised training data.  Further, generative adversarial networks are shown to synthesize novel RIS designs not previously seen in literature. This article provides a synopsis of DL techniques for inverse RIS design and optimization to yield targeted EM response necessary for future wireless networks.

	\end{abstract}

	\section{Introduction}
	\label{sec:Introduciton}
	
	The emerging industrial use-cases of sixth-generation (6G) and beyond wireless networks are envisaged to include industrial automation, autonomous vehicles, and smart infrastructure. These applications require significant improvements in data capacity, system latency, and quality-of-service reliability over the current 5G networks. In this context, \textit{reconfigurable intelligent surface} (RIS) has been identified as a key enabling technology to program the \textit{smart radio environment} (SRE), increase link quality, and reduce the hardware complexity \cite{hodge2019reconfigurable}. 
    
    The RIS is made up of a \textit{metasurface} (MTS) - a two-dimensional (2-D) reconfigurable electromagnetic (EM) layer composed of a large periodic array of subwavelength scattering elements (meta-atoms) with specially designed spatial features. Compared to electrically large arrays, the nearly passive meta-atoms offer lower cost and power consumption. The radio-frequency (RF) MTS performs customized transformations, such as beamforming, on a reflected incident wave through modified surface boundary conditions using Huygens’ principle. 
    The arrangement and subwavelength structure of each meta-atom and, in turn, the array of space- and time-varying meta-atoms determine MTS aperture field distribution and control the direction and strength of reflected signal \cite{hodge2019joint}. 
    
\begin{figure*}[t]
		\centering
		{\includegraphics[width=2.0\columnwidth]{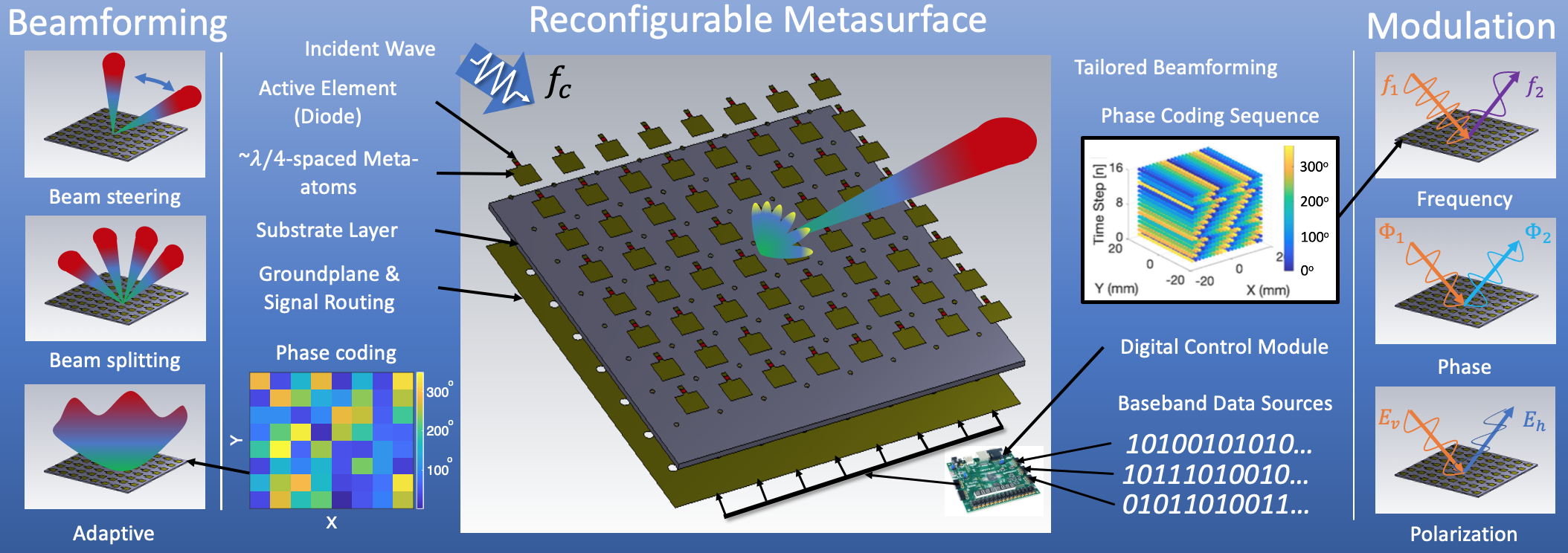}}
		\caption{The RIS architecture (center) operating at carrier frequency $f_c$ for wireless communication networks comprises meta-atoms located at below carrier wavelength ($\lambda$) spacing. It acts as both an endpoint transceiver and a relay. The RIS enables various beamforming functions (left column) including beam steering, splitting, and adaptive beamforming for customized radiation patterns by manipulating the phase coding of constituent meta-atoms. It is also capable of directly modulating (right column) the surface in frequency, phase, and polarization.
		}	
		\label{fig_SD_CE}
	\end{figure*}

    In a wireless link, the RIS functions as either an electrically large antenna array at the endpoints or as an amplify-and-forward relay (Fig. \ref{fig_SD_CE}). By actively controlling and optimizing the amplitude/phase of each meta-atom across the aperture, the RIS maximizes the receive signal-to-noise ratio and provides adaptive beamforming to coherently focus the reflected signal on the receiver. 
    Each scattering element typically includes an active tuning element, such as a varactor or PIN diode, whose bias voltage is software-controlled to change the EM response of the surface. The bias voltage for each meta-atom is pre-computed and modulated by a digital control module employing a field programmable gate array (FPGA) \cite{hodge2019reconfigurable}. Each meta-atom is controlled by tuning its EM properties (susceptibility or impedance) which affects the spectral response of the reflected signal. This aids in producing tailored radiation patterns for diverse functions, such as beam steering, anomalous reflection, focusing, beam splitting, absorption, and direct modulation of the reflected signal.  

However, the design and optimization of RIS hardware at the physical layer remains a formidable challenge. 
 In general, canonical structures such as v-antennas, loaded-dipoles, split-ring resonators, are used to fabricate RIS. However, meta-atoms based on these geometries usually fall short of desired performance, particularly when anisotropic, broadband, and/or wide-angle responses are required. 
 Designing a user-defined, arbitrary wave-front RIS or \textit{metagrating} is a challenging, labor-intensive, and long process. IN general, a new MTS design entails numerous rounds of manual tuning and full-wave simulations that iteratively solve Maxwell’s equations until a locally optimized design is achieved \cite{liu2018generative}. Initial designs are typically based on physical instincts and intuitive arguments. However, the final geometric structure and material characteristics are attained through iterative analyses.

	Recently, machine/deep learning (ML/DL) techniques have shown unprecedented performance in problems where it is difficult to develop an accurate mathematical model for feature representation. These methods are now also transforming the above-mentioned tedious approaches to design RIS and EM devices. Note that this is different than using DL to perform signal processing function in RIS-aided communications (see, e.g., \cite{elbir2020survey} for a survey). This paper provides an overview of recent developments in using ML/DL for designing the physical layer of RIS.

\section{Inverse RIS Design}
\label{sec:ris_design}
\begin{figure}[t]
    \centering
    \includegraphics[width=01.0 \columnwidth]{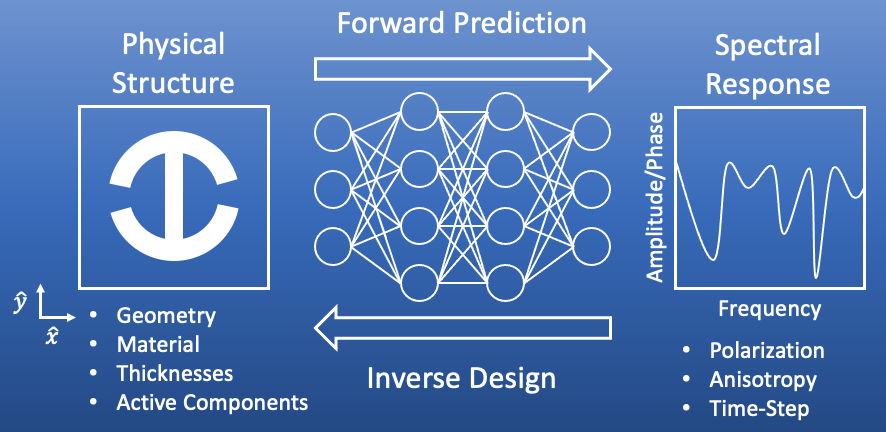}
    \caption{In inverse RIS design, ML algorithms learn and generalize complex EM relationships between the physical RIS structure (left column) and spectral response (right column) through training data. 
    }
    \label{fig:MLexample}
    \end{figure}

	Communications-based analysis of RIS without physics-based EM-compliant models is a major limitation of current research. 
	Until recently, prior works did not consider such realistic RIS implementations. As the parameter spaces of meta-atom geometry and constituent materials has grown, the conventional approaches to achieve the targeted EM response have become more tedious. In this context, deep learning models have demonstrated the ability to implicitly learn Maxwell's equations from training data within a constrained design space. The ML techniques have witnessed increased use in research to create surrogate models for MTS performance prediction, inverse design, and optimization. For an inverse MTS design problem, the input is an arbitrary design spectrum and the network finds or synthesizes a geometry to closely approximate the desired spectral response (Fig.~\ref{fig:MLexample}). 
	
	
	Major benefits of DL-based RIS design for wireless communications include:
	\begin{itemize}
    \item \textit{EM-based surrogate models}: DL constructs a nonlinear mapping between the raw input data (meta-atom design) and the desired output to approximate the MTS response.
    \item \textit{Inverse design}: Deep generative models are utilized to learn geometric features from training data and generate new meta-atom designs to achieve the spectral response.
    \item \textit{Diverse EM surface representations}: DL-based MTS design admits flexible design representation. The input could be either vectors of discrete parameters describing the geometry, material, frequency, and angular design parameters or pixelated images to represent the geometry or phases of the meta-atom design. Whereas a fully-connected neural network is well-suited to process the simple designs specified by the former representation, a convolutional networks handle images appropriately to yield more complex MTS geometries.
    \end{itemize}

\begin{table*}[t]
\centering
\caption{State-of-the-art on RIS inverse design}
\label{tab:my-table}
\begin{tabular}{||
>{\columncolor[HTML]{CBCEFB}}p{0.10\linewidth} ||
>{\columncolor[HTML]{EFEFEF}}p{0.09\linewidth} |
>{\columncolor[HTML]{CBCEFB}}p{0.04\linewidth} |
>{\columncolor[HTML]{EFEFEF}}p{0.06\linewidth} |
>{\columncolor[HTML]{CBCEFB}}p{0.31\linewidth} |
>{\columncolor[HTML]{EFEFEF}}p{0.23\linewidth} ||}
\hline
  \textbf{Algorithm} &  
  \textbf{Frequency} &
  \textbf{MTS layers} &
  \textbf{Data} &
  \textbf{Key features} &
  \textbf{Drawbacks} \\ \hline
\multicolumn{6}{|c|}{\textbf{Evolutionary optimization techniques}} \\ \hline 
GA \cite{campbell2019review}        &  $15$-$45$ GHz & $1$ & Parameter Vector & Pixelized meta-atoms with discrete input design space when a contiguous structure is not required & Optimization from scratch for each design; output structures may be too complex to fabricate \\ \hline
PSO \cite{campbell2019review}       & $9.5$-$12$ GHz & $1$  & Binary Matrix (2-D) & Swarm-based GO technique for pixelized meta-atom design; outperforms GA for various EM designs & Optimization from scratch for each design with parameter tuning \\ \hline
ACO \cite{campbell2019review}       & $1$-$4.5$ GHz & $3$ & Binary Matrix (3-D) & MTS, including 3-D structures and wire grid arrays, with discrete design space and a contiguous structure & Optimization from scratch for each design; output structures may be too complex to fabricate \\ \hline
\multicolumn{6}{|c|}{\textbf{Learning methods}} \\ \hline 
ANN \cite{peurifoy2018nanophotonic} & $375$-$749$ THz & $1$ & Parameter Vector & Performance prediction, inverse design, and optimization of nanophotonic particles & Limited design variables; applicable to only spherical dielectric nanoparticles \\ \hline
ANN \cite{inampudi2018neural}       & $193$ THz       & $1$ & Parameter Vector & Performance prediction and inverse design of metagratings & Limited set of parametric inputs; significant training overload  \\ \hline
DNN \cite{ma2018deep}               & $30$-$80$ THz   & $2$ &  Parameter Vector & Inverse design of chiral and multi-layer MTS & Design-specific architecture; 
limited design space \\ \hline
CNN \cite{zhang2019machine}         & $10$ GHz        & $1$ & Binary Matrix (2-D) & Anisotropic digital coding MTS; PSO for beamforming & Significant training overload  \\ \hline
CNN \cite{shan2020coding}           & $9.37$ GHz      & $1$ & Binary Matrix (2-D) & Hybrid CNN-GA for space-time modulation of programmable MTS; multi-beam steering & Binary phase coding limits beamforming performance; limited tunability \\ \hline
cDC-GAN \cite{liu2018generative}    & $170$-$600$ THz & $1$ & Image Matrix (2-D) & Generative inverse design of transmission MTS & Significant training overload; limited to single layer designs and passive structures \\ \hline
cDC-GAN \cite{hodge2019rf}          & $1$-$30$ GHz    & $1$ & Image (2-D) & Reflective RF MTS; training set with published meta-atom structures to improve learning  & Limited to single layer; post-processing required \\ \hline
cDC-GAN \cite{hodge2019joint}       & $5$-$25$ GHz    & $2$ & Image (3-D) & Multi-layer MTS; RGB-style matrix to represent multiple layers & No active elements; additional validation required \\ \hline
cDC-GAN \cite{hodge2019multi}       & $5$-$25$ GHz    & $3$ & Image (3-D) & Federated learning for multi-layer design & Significant training overload \\ \hline
cDC-VAE \cite{ma2019probabilistic}      & $40$-$100$ THz  & 1 & Image (2-D) & Anisotropic MTS; encodes input into low-dimensional latent space & 
Significant training overload; post-processing required
\\ \hline
TO-GAN \cite{jiang2019free}             & $231$-$600$ THz & 1 & Image (2-D) & Free-form diffractive metagrating design for select wavelength-deflection angle pairs with topology refinement & Additional optimization required 
\\ \hline
GLOnet \cite{jiang2019global}       & $231$-$500$ THz & 1 & Image (2-D) & 
Dielectric MTS design without training sets & Limited to single objective optimization; requires solving Maxwell's equations inside training loop \\ \hline
\end{tabular}
\end{table*}\normalsize

Table~\ref{tab:my-table} summarizes prior works on various techniques for RIS inverse design. The non-DL methods typically comprise of several evolutionary optimization algorithms as listed below.
	\subsubsection{Genetic algorithm (GA)}	
	This is an iterative global optimization (GO) algorithm that has been used extensively in the design of pixelated coded MTS designs. GA is a nature-inspired algorithm that uses binary strings (chromosomes) to represent candidate designs \cite{campbell2019review}. During the optimization, the GA selects the best subset of design candidates from the previous generation to serve as starting points for mutation and crossover in the next design iteration. Recent GA applications include coding MTS \cite{campbell2019review} which demonstrates channel response modification, efficient polarization conversion, and phase-graded beam steering.
	
	\subsubsection{Particle swarm optimization (PSO)}
	A popular stochastic evolutionary computation technique, PSO is inspired by the movement and intelligence of swarms. Recently, it has been employed for shaping EM waves using pixelized coded metasurfaces \cite{campbell2019review}. The design procedure using PSO is tied to a full-wave EM solver and completely automatic. The software yields both microscopic meta-atom designs and the macroscopic aperture coding matrix. By changing the reflection phase difference between cells, this approach has produced designs of functional metasurfaces with circularly- and elliptically-shaped radiation beams and multi-beam patterns. 
	Similar efforts have used a simulated annealing algorithm for the design and optimization of a broadband diffusion MTS using anisotropic elements for scattering reduction. In \cite{zhang2019machine}, binary PSO (BPSO) was used to automate the macroscropic layout of both passive and active aperture to realize user-defined dual-beam scattering radiation patterns. 

	
	\subsubsection{Ant colony optimization (ACO)} This is another swarm-based algorithm inspired by \textit{stigmergy} in ant colonies in order to search for optimal solutions to graph-based problems \cite{campbell2019review}. Here, a number of \textit{artificial ants} build solutions to an optimization problem and exchange information on their quality using a cooperation scheme similar to that utilized by real ants. In \cite{campbell2019review}, inverse MTS design is performed based on multi-objective lazy ACO (MOLACO) to synthesize 3-D nano-antenna geometries with low-loss transmission performance and broad phase tunability. The ACO is generally most useful for a discrete input design space and when a contiguous structure is required.
	
	\section{DL-Based Inverse Design and Optimization}
	The computational power and time required for evolutionary optimization algorithms grow exponentially with the number of design parameters. This is mitigated by DL-based inverse design for RIS. Prior works have employed a variety of network structures and algorithms based on the availability of data, RIS topology, and desired EM spectral response.
	\subsection{Artificial Neural Network (ANN)}
	\label{sec:ANNs}
    The ANNs were first used to approximate light scattering by multi-layer nanoparticles (meta-atoms) \cite{peurifoy2018nanophotonic}. Similar to MTS, nanophotonic particles derive their frequency response from physical structure and the size constituent scatterers. Then, \cite{inampudi2018neural} used a similar technique for metagratings. 
	The primary application of ANNs in MTS design is performance approximation. The feedforward ANN is trained to be a high-fidelity surrogate model for performance prediction. Using training data consisting of meta-atom physical design parameters as inputs and frequency response as labels, the ANN is trained to approximate a complex physics simulation (such as finite-element method (FEM), method of moment (MoM), or finite-difference time-domain (FDTD) simulation). Through the training data, the 
	ANN learns to map the scattering function of the meta-atom into a continuous, higher-order space where the derivative is found analytically through propagation. In \cite{peurifoy2018nanophotonic}, a trained ANN simulated spectral responses orders of magnitude faster than conventional full-wave simulations. 
	The results suggest that the ANN was not simply fitting the data, but rather discovered the underlying structure of input-to-output mapping to generalize the physics of the systems with the training set and solve problems not yet encountered. 
	
	A significant drawback of this approach is that the inputs are limited to the thicknesses of the meta-atom layers with fixed materials. This results in a lack of generalizability for the ANN that vastly limits the possible meta-atom design structures. While fixing the input parameters reduces the complexity of the ANN architecture, it limits the design space and optimal designs. Another drawback of this approach is that \cite{peurifoy2018nanophotonic} required $50,000$ examples using conventional simulation methods to generate training data. However, unlike evolutionary optimization methods such as GA or PSO, simulation of the training dataset is an upfront fixed cost because it only needs to be simulated once and is then leveraged for other designs. Additionally, the simulations for training data generation are highly parallelized unlike serial optimization techniques.

	Once trained, \cite{peurifoy2018nanophotonic} shows that the ANN solves inverse design problems more quickly than than its numerical counterparts because the gradient is found analytically, through back propagation, rather than numerically. Similar to inverse design, the ANN also optimizes for a desired property by altering the cost function used for the design without training the ANN. Their results that the ANN performs inverse design and optimization more accurately than traditional numerical nonlinear optimization techniques.
	
	\begin{figure*}[t]
		\centering
		\includegraphics[width=2.0\columnwidth]{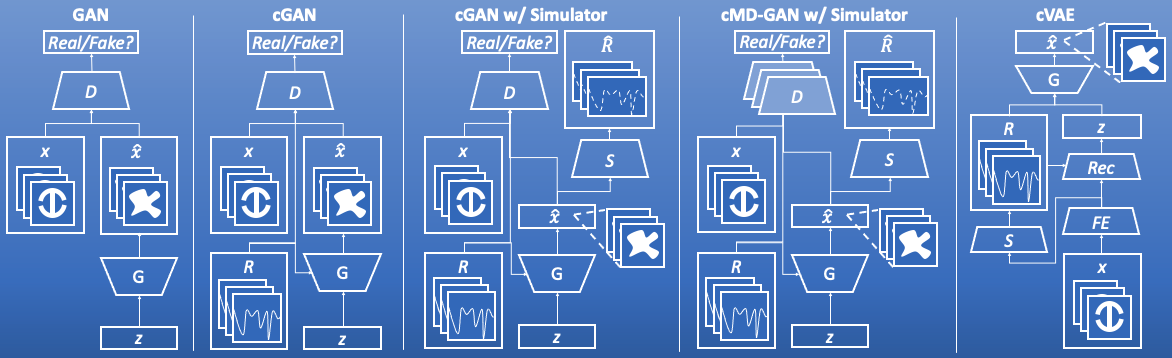}
		\caption{DGM architectures for RIS inverse design. The conventional GAN (left) lacks spectral information of the RIS structure $x$. The latent variable $z$ is fed to the generator $G$ to yield an estimated meta-atom structure $\hat{x}$. The discriminator $D$ then makes a decision if $\hat{x}$ is a valid design. The cGAN (second from left), conditioned by the reflection spectra $R$, shows improved performance. A simulator neural network may also be added to cGAN (center) to accelerate training and also predict the performance $\hat{R}$ of generated meta-atoms. The cMD-GAN (second from right) comprises of multiple discriminators, one for each layer. The cVAE (right) consists of an encoder-decoder network structure, where a feature extractor (FE) coupled with the recognition (Rec) network serves as the encoder to map the meta-atom structure to a lower-dimensional latent variable space. 
		The generation model is a reconstruction (decoder) network. 
		}
		\label{fig:netStructures}
	\end{figure*}
	
	\subsection{Deep Neural Networks (DNN)}
	To model more complex meta-atom structures and increase performance prediction accuracy, DL has been applied to the on-demand design of chiral (a form of anisotropy) MTS \cite{ma2018deep}. Here, DNN - an ANN comprised of many hidden layers to significantly expand learning and generalization ability - was employed to automatically design and optimize 3-D chiral MTS with strong anisotropic spectra at predetermined wavelengths. The network comprised two bidirectional networks that were constructed using partial stacking technique. This study limited the input design space (and hence the structures obtained) and predicted the reflection spectral response at $201$ discrete frequency points for two orthogonal polarization and the cross-polarization coupling term resulting in a $3$-by-$201$ spectral output vector. 
	Full-wave simulation was used to generate the training data set for $30,000$ example meta-atoms. The DNN achieved high efficiency and high-accuracy for performance prediction and inverse design for anisotropic MTS, where the meta-atom design space is limited.
	
	\subsection{Convolutional Neural Networks (CNN)}
	\label{sec:CNNs}
	To improve on the lack of generalization and increase performance prediction accuracy, CNNs are used to design anisotropic digital coding metasurfaces. CNNs are a class of ANNs that use convolution functions to learn hierarchical patterns within data. These models learn generalized patterns across many spatial scales from their input data and are widely used on image data. In \cite{zhang2019machine}, a CNN predicted the reflection phase response of binary coded meta-atoms where each meta-atom contains $16$-by-$16$ square sub-pixels and is mirrored with two-fold symmetry. 
	The results show an accuracy of $90.05\%$ of phase responses with $2^{\circ}$ error in the $360^{\circ}$ phase. A drawback of this binary coding approach is that a $16$-by-$16$ pixel meta-atom has $2^{16}$ potential design combinations. This study generated training data by simulating randomized pixel matrices. However, it was fundamentally inefficient in an analogous manner to GA because the training data is essentially random and does not contain the knowledge of canonical structures in the training data set. 
	A significant CNN advantage is that the meta-atom shape is directly input into the network rather than shape-specific design parameters. The convolutional filters allow the CNN to learn the physical structure that leads to given EM response, leading to a broader applicability of the model.
	
	In \cite{shan2020coding}, the element phases of a reconfigurable MTS were computed by a 11-layer CNN for multiple beam steering applications. The input was the parameter vector representing the target beam pattern and the output was a matrix that carried the 1-bit codes for a programmable $2304$-element MTS. This technique to obtain the phase matrices reduced the time for producing almost similar beam patterns using conventional methods to a few milliseconds.

	\subsection{Deep Generative Models (DGMs)}
	Generative models are unsupervised or semi-supervised learning models that infer a function to describe hidden structure from unlabeled data. Their functions include clustering, density estimation, feature learning, and dimension reduction. Whereas discriminative networks capture the relationship between meta-atom geometry and spectral response from a training set, DGMs focus on learning the properties of meta-atom geometry distributions \cite{liu2018generative, hodge2019rf, jiang2019free, jiang2019global}. Major classes of DGMs (Fig.~\ref{fig:netStructures}) applied to MTS inverse design are as follows.
	\subsubsection{Generative adversarial networks (GANs)}
	\label{sec:GANs}
	In a GAN system, two ANNs compete to improve each of their models: the generative network learns to create inputs indistinguishable from the training data while the discriminative network learns to identify true data from the output of the generative network. 
	GANs are promising for low-cost MTS design with complex frequency and polarization dependent scattering responses. In \cite{liu2018generative}, an input set of user-defined EM spectra is fed to GAN that generates candidate patterns to match the on-demand spectra with high fidelity. Here, DNNs are employed to approximate the spectra of the MTS and perform inverse design by generating meta-atom structures that yield user-defined input spectra. Once the model is trained, extensive parameter scans and trial-and-error procedures are bypassed. This conditional deep convolutional GAN (cDC-GAN) architecture uses three interconnected CNNs: generator, discriminator, and simulator. The simulator is a pretrained network that serves as a surrogate model for fast spectral performance prediction. 
	The conditional generator networks accepts the desired spectral response and a latent noise vector to output potential meta-atom designs. The discriminator serves to train the generator by evaluating the distance of the distributions between the geometric patterns from training data and generator. At the end of successful training, discriminator is unable to distinguish batches from generator and training set. This approach is shown to exhibit high accuracy in inverse design of meta-atoms.

	In \cite{hodge2019rf}, a deep convolutional GAN (DC-GAN) is employed to generate anisotropic RF meta-atom designs. Using a small set of simulated spectra, the network learned the relationship between the physical structure of meta-atoms and their reflection spectra for vertical and horizontal polarizations. The DC-GANs generated meta-atom structures that resembled design features in the training data. To speed up training, the network was fed with parametric variations of twelve published meta-atom designs to a full-wave EM simulator. Starting out with parametric variations of canonical meta-atoms scatterers, the network picked up more efficiently than it would have from training with responses of randomized pixel data.
	
\begin{figure}[t]
		\centering
		\includegraphics[width=\columnwidth]{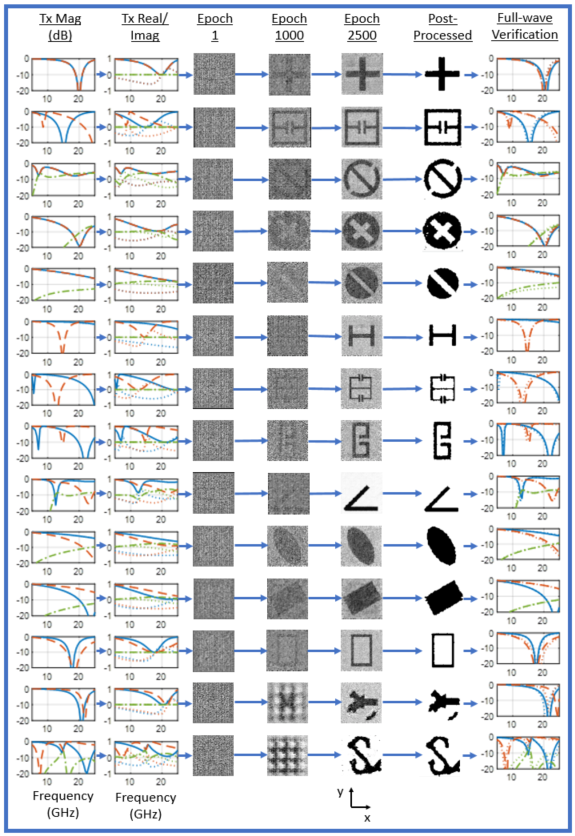}
		\caption{Meta-atom structures generated using DC-GAN \cite{hodge2019joint}. The first twelve rows show the ability of the DC-GAN to regenerate canonical structures from the training data set. The last two rows show the ability of the DC-GAN to generate new meta-atom geometries, exhibiting spatial features similar to those in the training data set.  
		}
		\label{fig:numExpResults}
	\end{figure}

\begin{figure}[t]
		\centering
		\includegraphics[width=0.9\columnwidth]{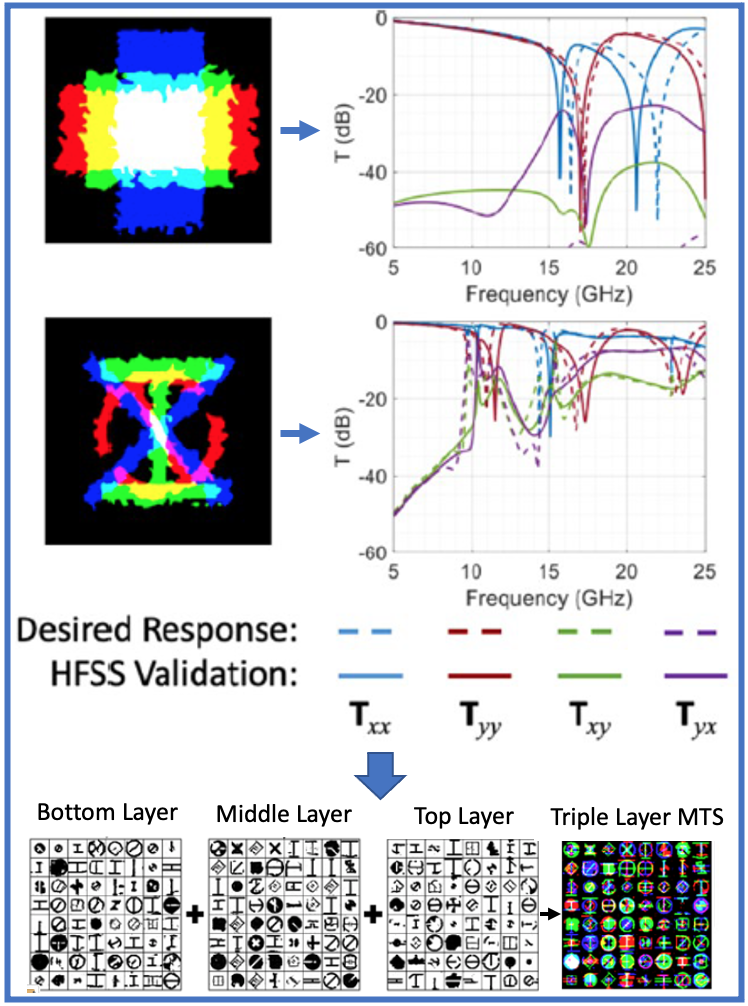}
		\caption{
		Two three-layer meta-atom designs (upper left) generated from cMD-GAN in \cite{hodge2019multi}. The top (red), middle (green), and bottom (blue) layers are metallic traces separated by dielectric spacers. The image matrices are post-processed to remove the background noise. The upper right shows desired input RF transmission response vectors (dashed lines) and the full-wave verification using the Ansys HFSS finite element method solver (solid lines) of generated design for $5$-$25$ GHz when illuminated by a plane wave at normal incidence. The blue and maroon lines represent the respective $x$ and $y$ co-polarized transmission responses. Similarly, the green and purple lines represent the cross-polarized responses for $x$ and $y$ polarized signals. The bottom row shows a composite of each layer of generated meta-atoms for a three-layer RIS.
		}
		\label{fig:resultsMDgan}
	\end{figure}

	The design approach in \cite{hodge2019joint} introduced the cDC-GAN-based for jointly designing several layers of tensorial RIS. It represented three RIS layers with a $64\times 64\times 3$ red-green-blue (RGB) image matrix. 
	The co-polarization and cross-polarization transmission responses of the resulting meta-atom designs (Fig.~\ref{fig:numExpResults}) differed from EM simulators by less than a dB. One of the most exciting features of cDC-GAN is its ability to discover new geometries not previously found in the literature. This suggests that the model implicitly learned the physical relationships of Maxwell's equations rather than simply interpolating from past designs. Building on this techniques, the federated learning approach in \cite{hodge2019multi} employed a conditional multi-discriminator distributed GAN (cMD-GAN) (see Fig.~\ref{fig:netStructures}) for multi-layer RF MTS discovery (Fig.~\ref{fig:resultsMDgan}).
	
	
	\subsubsection{Conditional variational autoencoder (cVAE)}
	As an alternative to GAN approaches, \cite{ma2019probabilistic} presents a probabilistic DGM that solves both forward and inverse problems at the same time. It is trained in an end‐to‐end manner and uses a deep convolution cVAE (cDC-VAE) architecture (Fig.~\ref{fig:netStructures}) comprising an encoder-decoder network structure. The encoder maps the meta-atom structure to a multivariate Gaussian distribution in the latent space and the conditional decoder network inputs the reflection spectra and latent variable to generate meta-atom designs (Fig.~\ref{fig:netStructures}).
	
	In \cite{ma2019probabilistic}, the RIS inverse design is modeled in a probabilistic generative manner to investigate the complex structure–performance relationship in an interpretable way and solve the one-to-many mapping issue that is intractable in deterministic models. It developed a semi-supervised learning strategy that allows the model to utilize unlabeled data in addition to labeled data in an end-to-end training. The RIS design and spectral response are encoded into a low-dimensional latent space with a predefined prior distribution, from which the latent variables are sampled. 
	
	    
	\subsubsection{Global topology optimization networks (GLOnets)}
	Recently, GANs utilized to learn structural features of topology-optimized (TO) metagratings for inverse design \cite{jiang2019free, jiang2019global}. TO is a method of optimizing a material layout or an array of pixels to maximize system performance given a set of constraints and boundary conditions. Unlike other approaches, simulation overload for TO does not increase with the number of RIS units. In \cite{jiang2019free}, free-form diffractive metagratings were designed using TO-GAN. Here, DGMs were trained from images of periodic, TO metagratings to produce efficient scattering structures with the desired performance over a broad range of frequencies and angles. The network employed $5,000$ training examples for each angle. However, the performance of the best structures was not robust and additional refinement was needed to meet the desired performance.
	In \cite{jiang2019global}, dielectric metasurfaces optimization was performed using a physics-informed cGAN. Global optimization-based generative networks (GLOnets) are able to search the design space for the global optimum design. Unlike other GAN approaches, GLOnets seek to fit a narrow-peaked function centered on the optimal solution without a training set. The GLOnet generates a distribution of meta-atoms to samples the global design space and then shifts the distribution toward a more optimal design. Training requires computing forward and adjoint EM simulations of output structures using backpropagation. In this work, GLOnets are shown to be successful and computationally efficient global TO for MTS and metagratings.

	\section{Challenges and Future Research Directions}
	\label{sec:pathForward}
	The techniques for RIS inverse design are constantly evolving. As mentioned below, reduction of design time and achieving full EM-compliance remains a major challenge.
	
	\subsubsection{Hybrid physics-based models}
	New approaches are needed to increase the computational efficiency and reduce the amount of training required for DL-based RIS design. Hybrid models, where training set is supplemented by physics-based analytical models, reduce the amount of required training data and increase learning efficiency. Analytical RF circuit-based models are available to predict the performance of several canonical meta-atom designs. To speed up the training data generation, these analytical circuit-based models could be used to supplement the training data set and reduce iterations of time-consuming full-wave EM simulations. 
	It may also be feasible to create innovative DL design and optimization architectures that utilize physics-based analytical models within the ANN architecture. 
	
	\subsubsection{Other learning techniques}
	Transfer learning (TL) is a technique for expediting and improving the learning of a new task by using a previously trained neural network weights and bias as the initialization for the new ANN. Since all ANNs for meta-atom performance prediction and inverse design are implicitly learning Maxwell's equations, it is sensible that a network trained for one meta-atom design or frequency band is scaled and transferred to a related design. Deep Q-networks (DQNs) have also been studied to increase the efficiency of MTS holograms and automated multi-layer RIS design.
	
	\subsubsection{Improved data representation}
	More complex input data structures and representations are increasingly studied for DL-based RIS. While this article focused on discrete input parameters and image data structures are RIS design representations,  graphical and sequential data structures have recently been proposed as alternatives. The graphical model has been used to represent EM systems with near-field coupling (as in coupled resonators). In this arrangement, graph nodes contain resonator attributes, such as material, geometry, and location, and graph nodes represent the near-field coupling factors. These graphical data structures are processed using graphical neural networks (GNNs). 
	Similarly, sequential data structures are useful for representing time-sequence data in dynamic EM systems (as in RIS filters) and are learned using recurrent neural networks (RNNs). 
	
	
	\section{Summary}
	\label{sec:Conc}
	We surveyed DL-based techniques for designing RIS hardware to be deployed for future wireless communications. When the design space and scale of the RIS arrays increases, learning-based architectures outperform evolutionary optimization techniques for both surrogate performance modeling and inverse design. The DL inverse design is flexible in admitting a variety of RIS unit structures. The DGMs are the most useful because of their ability to generate new designs not previously seen in the published literature. While active research and techniques in this area are still evolving, DL is a promising solution for the inverse design of RIS.
	
	
	\balance
	\bibliographystyle{IEEEtran}
	\bibliography{main}

\begin{thebibliography}{10}
\providecommand{\url}[1]{#1}
\csname url@samestyle\endcsname
\providecommand{\newblock}{\relax}
\providecommand{\bibinfo}[2]{#2}
\providecommand{\BIBentrySTDinterwordspacing}{\spaceskip=0pt\relax}
\providecommand{\BIBentryALTinterwordstretchfactor}{4}
\providecommand{\BIBentryALTinterwordspacing}{\spaceskip=\fontdimen2\font plus
\BIBentryALTinterwordstretchfactor\fontdimen3\font minus
  \fontdimen4\font\relax}
\providecommand{\BIBforeignlanguage}[2]{{%
\expandafter\ifx\csname l@#1\endcsname\relax
\typeout{** WARNING: IEEEtran.bst: No hyphenation pattern has been}%
\typeout{** loaded for the language `#1'. Using the pattern for}%
\typeout{** the default language instead.}%
\else
\language=\csname l@#1\endcsname
\fi
#2}}
\providecommand{\BIBdecl}{\relax}
\BIBdecl

\bibitem{hodge2019reconfigurable}
J.~A. Hodge, K.~V. Mishra, and A.~I. Zaghloul, ``Reconfigurable metasurfaces
  for index modulation in {5G} wireless communications,'' in \emph{IEEE
  International Applied Computational Electromagnetics Society Symposium},
  2019, pp. 1--2.

\bibitem{hodge2019joint}
------, ``Joint multi-layer {GAN}-based design of tensorial {RF}
  metasurfaces,'' in \emph{IEEE International Workshop on Machine Learning for
  Signal Processing}, 2019, pp. 1--6.

\bibitem{liu2018generative}
Z.~Liu, D.~Zhu, S.~P. Rodrigues, K.-T. Lee, and W.~Cai, ``Generative model for
  the inverse design of metasurfaces,'' \emph{Nano letters}, vol.~18, no.~10,
  pp. 6570--6576, 2018.

\bibitem{elbir2020survey}
A.~M. Elbir and K.~V. Mishra, ``A survey of deep learning architectures for
  intelligent reflecting surfaces,'' \emph{arXiv preprint arXiv:2009.02540},
  2020.

\bibitem{campbell2019review}
S.~D. Campbell, D.~Sell, R.~P. Jenkins, E.~B. Whiting, J.~A. Fan, and D.~H.
  Werner, ``Review of numerical optimization techniques for meta-device
  design,'' \emph{Optical Materials Express}, vol.~9, no.~4, pp. 1842--1863,
  2019.

\bibitem{peurifoy2018nanophotonic}
J.~Peurifoy, Y.~Shen, L.~Jing, Y.~Yang, F.~Cano-Renteria, B.~G. DeLacy, J.~D.
  Joannopoulos, M.~Tegmark, and M.~Solja{\v{c}}i{\'c}, ``Nanophotonic particle
  simulation and inverse design using artificial neural networks,''
  \emph{Science advances}, vol.~4, no.~6, p. eaar4206, 2018.

\bibitem{inampudi2018neural}
S.~Inampudi and H.~Mosallaei, ``Neural network based design of metagratings,''
  \emph{Applied Physics Letters}, vol. 112, no.~24, p. 241102, 2018.

\bibitem{ma2018deep}
W.~Ma, F.~Cheng, and Y.~Liu, ``Deep-learning-enabled on-demand design of chiral
  metamaterials,'' \emph{ACS nano}, vol.~12, no.~6, pp. 6326--6334, 2018.

\bibitem{zhang2019machine}
Q.~Zhang, C.~Liu, X.~Wan, L.~Zhang, S.~Liu, Y.~Yang, and T.~J. Cui,
  ``Machine-learning designs of anisotropic digital coding metasurfaces,''
  \emph{Advanced Theory and Simulations}, vol.~2, no.~2, p. 1800132, 2019.

\bibitem{shan2020coding}
T.~Shan, X.~Pan, M.~Li, S.~Xu, and F.~Yang, ``Coding programmable metasurfaces
  based on deep learning techniques,'' \emph{IEEE Journal on Emerging and
  Selected Topics in Circuits and Systems}, vol.~10, no.~1, pp. 114--125, 2020.

\bibitem{hodge2019rf}
J.~A. Hodge, K.~V. Mishra, and A.~I. Zaghloul, ``{RF} metasurface array design
  using deep convolutional generative adversarial networks,'' in \emph{IEEE
  International Symposium on Phased Array Systems and Technology}, 2019.

\bibitem{hodge2019multi}
------, ``Multi-discriminator distributed generative model for multi-layer {RF}
  metasurface discovery,'' in \emph{IEEE Global Conference on Signal and
  Information Processing}, 2019.

\bibitem{ma2019probabilistic}
W.~Ma, F.~Cheng, Y.~Xu, Q.~Wen, and Y.~Liu, ``Probabilistic representation and
  inverse design of metamaterials based on a deep generative model with
  semi-supervised learning strategy,'' \emph{Advanced Materials}, vol.~31,
  no.~35, p. 1901111, 2019.

\bibitem{jiang2019free}
J.~Jiang, D.~Sell, S.~Hoyer, J.~Hickey, J.~Yang, and J.~A. Fan, ``Free-form
  diffractive metagrating design based on generative adversarial networks,''
  \emph{ACS nano}, vol.~13, no.~8, pp. 8872--8878, 2019.

\bibitem{jiang2019global}
J.~Jiang and J.~A. Fan, ``Global optimization of dielectric metasurfaces using
  a physics-driven neural network,'' \emph{Nano letters}, vol.~19, no.~8, pp.
  5366--5372, 2019.

\end{thebibliography}
    
    \begin{IEEEbiographynophoto} 
		{John A. Hodge} (S'12-M'14) obtained M.S. in electrical engineering from Virginia Tech and a dual undergraduate degree in electrical \& computer engineering and physics from Duke University. Currently, John is a Ph.D. candidate in electrical engineering at Virginia Tech, under the guidance of Dr. Amir Zaghloul. From 2012 to 2014, he studied electromagnetics and antennas as a graduate research assistant at the U.S. Army Research Lab (ARL) in Adelphi, MD. John’s Ph.D. dissertation focuses on reconfigurable metasurface antennas for communications and radar applications. He is the recipient of several student paper awards. John is also a senior principal RF design engineer at Northrop Grumman in Baltimore, Maryland, where he is involved in the design of wideband phased arrays.
	\end{IEEEbiographynophoto} 
	\vskip -2\baselineskip plus -1fil
	\begin{IEEEbiographynophoto} 
		{Kumar Vijay Mishra} (S'08-M'15-SM'18) obtained a Ph.D. in electrical engineering and M.S. in mathematics from The University of Iowa in 2015, M.S. in electrical engineering from Colorado State University in 2012, B. Tech. \textit{summa cum laude} (Gold Medal, Honors) in electronics and communication engineering from the National Institute of Technology, Hamirpur, India in 2003. He is currently U. S. National Academies Harry Diamond Distinguished Fellow at the United States Army Research Laboratory; Technical Adviser to Hertzwell, Singapore; and honorary Research Fellow at SnT, University of Luxembourg. He is the recipient of several fellowships and best paper awards. He has been an Associate Editor (Radar Systems) of IEEE Transactions on Aerospace and Electronic Systems since 2020. 
	\end{IEEEbiographynophoto} 
	\vskip -2\baselineskip plus -1fil
	\begin{IEEEbiographynophoto} 
	{Amir I. Zaghloul} (S’68-M’73-SM’80-F’92-LF’11) received the B.Sc. degree (Hons.) from Cairo University, Egypt, in 1965, and the M.A.Sc. and Ph.D. degrees from the University of Waterloo, Waterloo, ON, Canada, in 1970 and 1973, respectively, all in electrical engineering. He is an ARL Fellow with the U.S. Army Research Laboratory (ARL), Adelphi, MD, USA, and a Research Professor with Virginia Tech, Blacksburg, VA, USA. Dr. Zaghloul is an IEEE Life Fellow, Fellow of URSI, Fellow of the Applied Computational Electromagnetics Society (ACES), and Associate Fellow of The American Institute of Aeronautics and Astronautics. He is Chair of Commission C of URSI. He received several research and patent awards, including the Exceptional Patent Award at COMSAT and the 1986 Wheeler Prize Award for Best Application Paper in the IEEE Transactions on Antennas and Propagation.
	\end{IEEEbiographynophoto}

\end{document}